%% file: 0-main.tex
\documentclass{article}
\usepackage{spconf,amsmath,graphicx}
\usepackage{amsfonts}
\usepackage{multirow}
\usepackage{array}
\usepackage{cite}
\usepackage{bibspacing}
\usepackage{subfig}
\usepackage{color}
\usepackage{graphicx} 
\usepackage{epstopdf}
\usepackage{makecell}
\usepackage{mathtools}

\usepackage{pifont}
\newcommand{\cmark}{\ding{51}}
\newcommand{\xmark}{\ding{55}}
\renewcommand\arraystretch{1.0}
\renewcommand\tabcolsep{2.0pt}
\newcommand{\tabincell}[2]{\begin{tabular}{@{}#1@{}}#2\end{tabular}}

\usepackage[normalem]{ulem}

\makeatletter
\def\bstctlcite{\@ifnextchar[{\@bstctlcite}{\@bstctlcite[@auxout]}}
\def\@bstctlcite[#1]#2{\@bsphack
  \@for\@citeb:=#2\do{%
    \edef\@citeb{\expandafter\@firstofone\@citeb}%
    \if@filesw\immediate\write\csname #1\endcsname{\string\citation{\@citeb}}\fi}%
  \@esphack}
\makeatother

\title{Enhancing Pre-trained ASR System Fine-tuning for Dysarthric Speech Recognition using Adversarial Data Augmentation}
\name{Huimeng Wang$^*$,Zengrui Jin$^*$,Mengzhe Geng,Shujie Hu,Guinan Li,Tianzi Wang,Haoning Xu,Xunying Liu}
\address{\textit{\{huimengwang, zrjin, mzgeng, sjhu, gnli, twang, hnxu, xyliu\}@se.cuhk.edu.hk} \\
The Chinese University of Hong Kong, Hong Kong SAR, China}

\begin{document}\bstctlcite{IEEEexample:BSTcontrol}
\setlength{\bibitemsep}{.02\baselineskip}

\ninept
\maketitle
\begin{abstract}
Automatic recognition of dysarthric speech remains a highly challenging task to date. 
Neuro-motor conditions and co-occurring physical disabilities create difficulty in large-scale data collection for ASR system development. 
Adapting SSL pre-trained ASR models to limited dysarthric speech via data-intensive parameter fine-tuning leads to poor generalization. 
To this end, this paper presents an extensive comparative study of various data augmentation approaches to improve the robustness of pre-trained ASR model fine-tuning to dysarthric speech. 
These include: a) conventional speaker-independent perturbation of impaired speech; b) speaker-dependent speed perturbation, or GAN-based adversarial perturbation of normal, control speech based on their time alignment against parallel dysarthric speech; c) novel Spectral basis GAN-based adversarial data augmentation operating on non-parallel data. 
Experiments conducted on the UASpeech corpus suggest GAN-based data augmentation consistently outperforms fine-tuned Wav2vec2.0 and HuBERT models using no data augmentation and 
speed perturbation across different data expansion operating points by statistically significant word error rate (WER) reductions up to 2.01\% and 0.96\% absolute (9.03\% and 4.63\% relative) respectively on the UASpeech test set of 16 dysarthric speakers.  
After cross-system outputs rescoring, the best system produced the lowest published WER of 16.53\% (46.47\% on very low intelligibility) on UASpeech. 
\end{abstract}

\let\thefootnote\relax\footnotetext{* Equal contribution was made between the first two authors.}
\vspace*{-1mm}
\section{Introduction}
\label{sec:intro}
\input{1-intro}
\vspace*{-2.5mm}
\section{Pre-trained ASR systems}
\vspace{-1mm}
\input{2-SSL_models}

\vspace*{-1.5mm}
\section{Conventional Data Augmentation}
\label{sec:conv_da}
\input{3-Conventional-DA}

\vspace{-2mm}
\section{Adversarial Data Augmentation}
\vspace{-1mm}
\input{4-Adversarial-DA}

\vspace*{-2mm}
\section{Experiments}
\label{sec:exp}
\input{5-Experiments}

\vspace*{-0.9mm}
\section{Conclusions}
\vspace{-2.5mm}
\input{6-Conclusions}

\vspace{-2.5mm}
\section{Acknowledgements}
\vspace{-3mm}
This research is supported by Hong Kong RGC GRF grant No. 14200021, 14200220, Innovation \& Technology Fund grant No. ITS/254/19 and ITS/218/21.

\newpage

\bibliographystyle{IEEEtran}
\bibliography{refs}

\end{document}

%% file: 1-intro.tex
\input{Figure/fig}
\vspace*{-1mm}
Despite the rapid progress of automatic speech recognition (ASR) technologies targeting normal speech, accurate recognition of pathological voice, for example, dysarthric speech, remains a challenging task\cite{inproceedings,liu21recent,xiong2020source,jin21_interspeech,yue22acoustic,geng2023fly} to data due to: a) the scarcity of such data; b) their large mismatch against normal speech; and c) large speaker level diversity.
The physical disabilities and mobility limitations associated with impaired speakers increase the difficulty of collecting large quantities of disordered speech for ASR system development. 
To this end, data augmentation techniques based on, for example, temporal or speed perturbation \cite{WSOLA,elasticspectraldistortion,ko15_interspeech,geng2020investigation}, GAN-based adversarial augmentation \cite{jin21_interspeech,harvill2021synthesis,DCGANs2018,prananta22_interspeech}, voice conversion \cite{vcJapan,huang21d_interspeech} and text-to-speech synthesis \cite{wang23qa_interspeech} have been developed and play a vital role in addressing the data scarcity issue for dysarthric speech recognition.

An alternative solution to address the above data sparsity issue is to use self-supervised learning (SSL) based speech foundation models  \cite{baevski2020wav2vec, chen2022wavlm, hsu2021hubert} pre-trained on large quantities of unlabelled data. These speech foundation models have been successfully applied to a variety of downstream tasks including, but not limited to speech recognition \cite{hsu2021hubert,chen2022wavlm,baevski2020wav2vec}, speech emotion recognition \cite{speechemotion} and speaker recognition \cite{vaessen2022fine}.
In contrast, limited prior researches have been conducted on disordered speech recognition using SSL pre-trained ASR models \cite{hernandez22_interspeech, lester2022investigatiing, baskar22b_interspeech,hu2023exploring,AVSSL,wang2023benefits}.
Prior researches in this area largely target English dysarthric speech, while producing mixed results on benchmark datasets represented by the UASpeech \cite{uaspeech2008} task. Among these, Wav2vec2.0 \cite{baevski2020wav2vec} and WavLM \cite{chen2022wavlm} models were studied in \cite{lester2022investigatiing}, where an average word error rate (WER) of 51.8\% was reported on the UASpeech test set of 16 dysarthric speakers.
Cross-lingual XLSR model \cite{conneau21_interspeech} based dysarthric speech recognition was studied in \cite{hernandez22_interspeech}, 
and produced average WERs of 62.0\% and 28.6\% on the very low and low intelligibility subsets of UASpeech.
Audio-visual pre-trained AV-HuBERT models \cite{shi2022learning} were employed in \cite{AVSSL} 
to produce a WER of 63.98\% on the very low intelligibility data of the UASpeech test data. 
A range of approaches to integrating SSL pre-trained models and their features into in-domain dysarthric speech trained ASR systems are explored in \cite{hu2023exploring} while producing an average WER of 22.83\% on UASpeech (52.53\% and 25.00\% on the very low and low intelligibility subsets). 

Efforts on applying SSL pre-trained speech models to dysarthric speech are confronted with the same data scarcity issue discussed above. Domain adapting SSL pre-trained ASR models containing a large number of model parameters to limited dysarthric speech via data-intensive fine-tuning rapidly leads to poor generalization. This issue is further exasperated when the limited training data provides insufficient coverage of words in the test data. For example, approximately 39\% of words in the benchmark UASpeech test set do not occur in the training data. In previous studies \cite{liu21recent,hu2023exploring,lester2022investigatiing}, this coverage issue was found to produce a large disparity in ASR performance between seen and unseen words in dysarthric speech and between impaired speakers with high and very low intelligibility. 

To this end, this paper presents an extensive comparative study over various data augmentation approaches to improve the robustness of pre-trained ASR model fine-tuning to scarce dysarthric speech. These include: a) the conventional speaker-independent perturbation of in-domain impaired speech; b) speaker-dependent speed perturbation, or GAN-based adversarial perturbation of normal, control speech based on their time alignment against parallel dysarthric speech; c) novel Spectral basis GAN based data augmentation that can operate on non-parallel data where the transformation between normal and dysarthric speakers’ data is implemented via the adversarial mapping between their respective time-invariant, content-independent spectral basis features. 

Experiments conducted on the UASpeech corpus suggest GAN-based data augmentation consistently outperforms fine-tuned Wav2vec2.0 and HuBERT models using no data augmentation and 
speed perturbation across different data expansion operating points by statistically significant word error rate (WER) reductions up to 2.01\% and 0.96\% absolute (9.03\% and 4.63\% relative) respectively on the UASpeech test set of 16 dysarthric speakers.  
After cross-system outputs rescoring, the best system produced the lowest published WER of 16.53\% (46.47\% and 16.76\% on very low and low intelligibility respectively) on the UASpeech task. 

The main contributions of this paper are summarized below:

\noindent{\bf 1)} To the best of our knowledge, this paper presents the first comparative study over various audio perturbation and adversarial data augmentation approaches for robust domain fine-tuning of SSL pre-trained ASR models for dysarthric speech recognition tasks. In contrast, prior researches on data augmentation for SSL pre-trained model fine-tuning were limited to normal speech domains \cite{huh2023comparison}, while adversarial data augmentation approaches were previously studied only in the context of non-SSL pre-trained ASR systems constructed using in-domain impaired speech only \cite{DCGANs2018,jin21_interspeech,harvill2021synthesis,jin2022personalized,jin23vae,baali23_interspeech,prananta22_interspeech}.

\noindent{\bf 2)} The proposed Spectral basis GAN model benefits from the disentanglement of time-invariant speaker-specific characteristics, such as an average reduction in speech volume and clarity found in impaired speakers,  from time variant temporal features that are more related to spoken contents. 
This novel approach broadens the application scope of GAN-based data augmentation approaches that traditionally require the use of parallel normal-dysarthric speech recordings. 

\noindent{\bf 3)} The final best-performing combined system featuring different forms of data augmentation techniques and pre-trained ASR models produced the lowest published WER of 16.53\% (46.47\% and 16.76\% on very low and low intelligibility respectively) on the UASpeech test set of 16 dysarthric speakers. 

%% file: Figure/fig.tex
\begin{figure*}[!ht]
\centering
\subfloat[] {
    \includegraphics[height=0.15\linewidth]{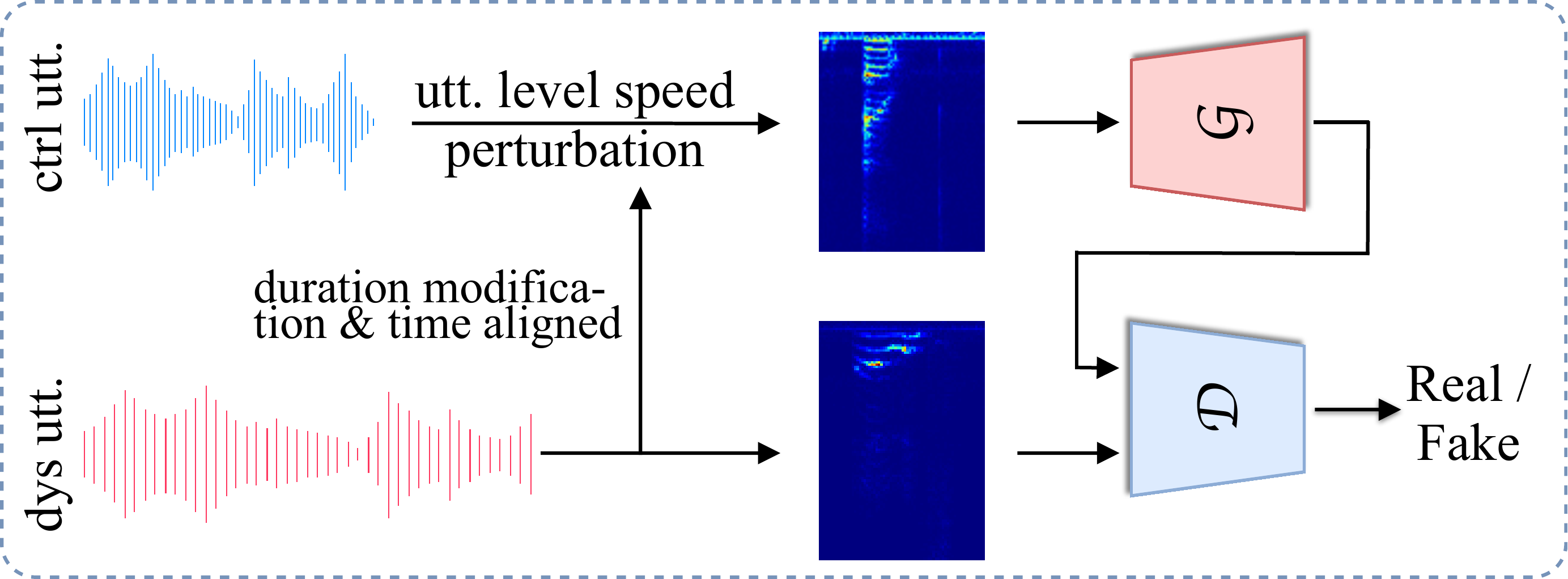}
    \label{fig:subfig:training}
}\hfill
\subfloat[] {
    \includegraphics[height=0.15\linewidth]{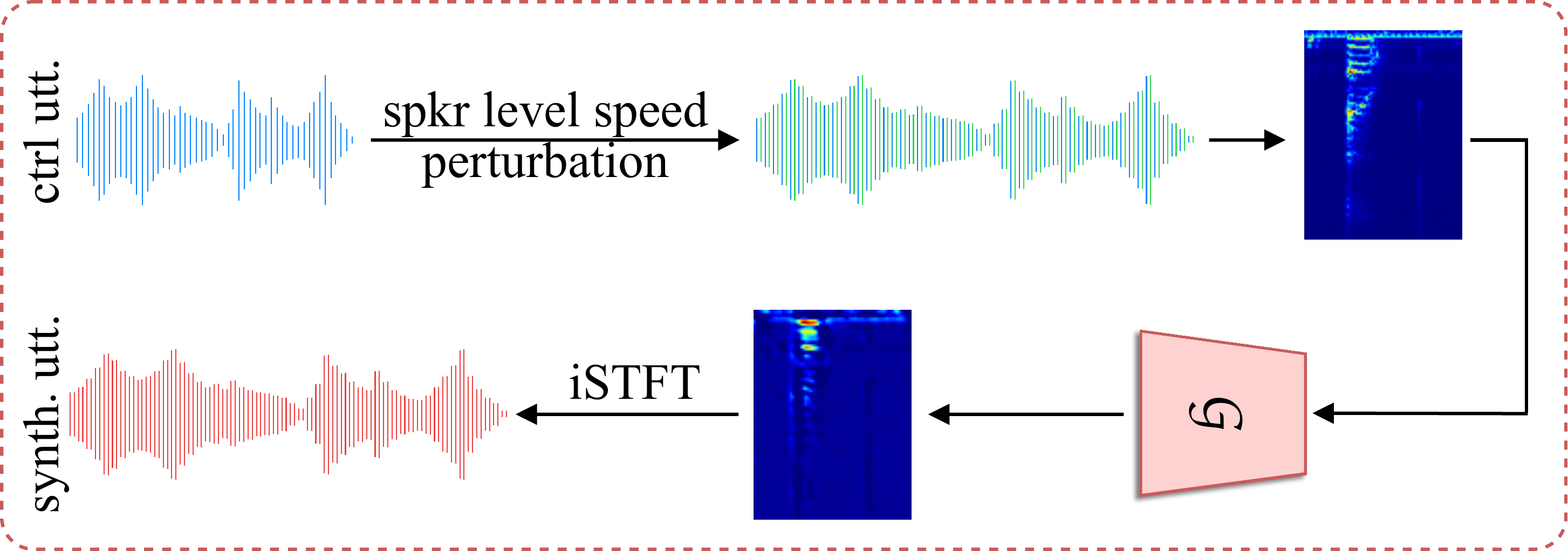}
    \label{fig:subfig:conversion}
}
    \vspace{-4mm}
\subfloat[] {
    \includegraphics[height=0.203\linewidth]{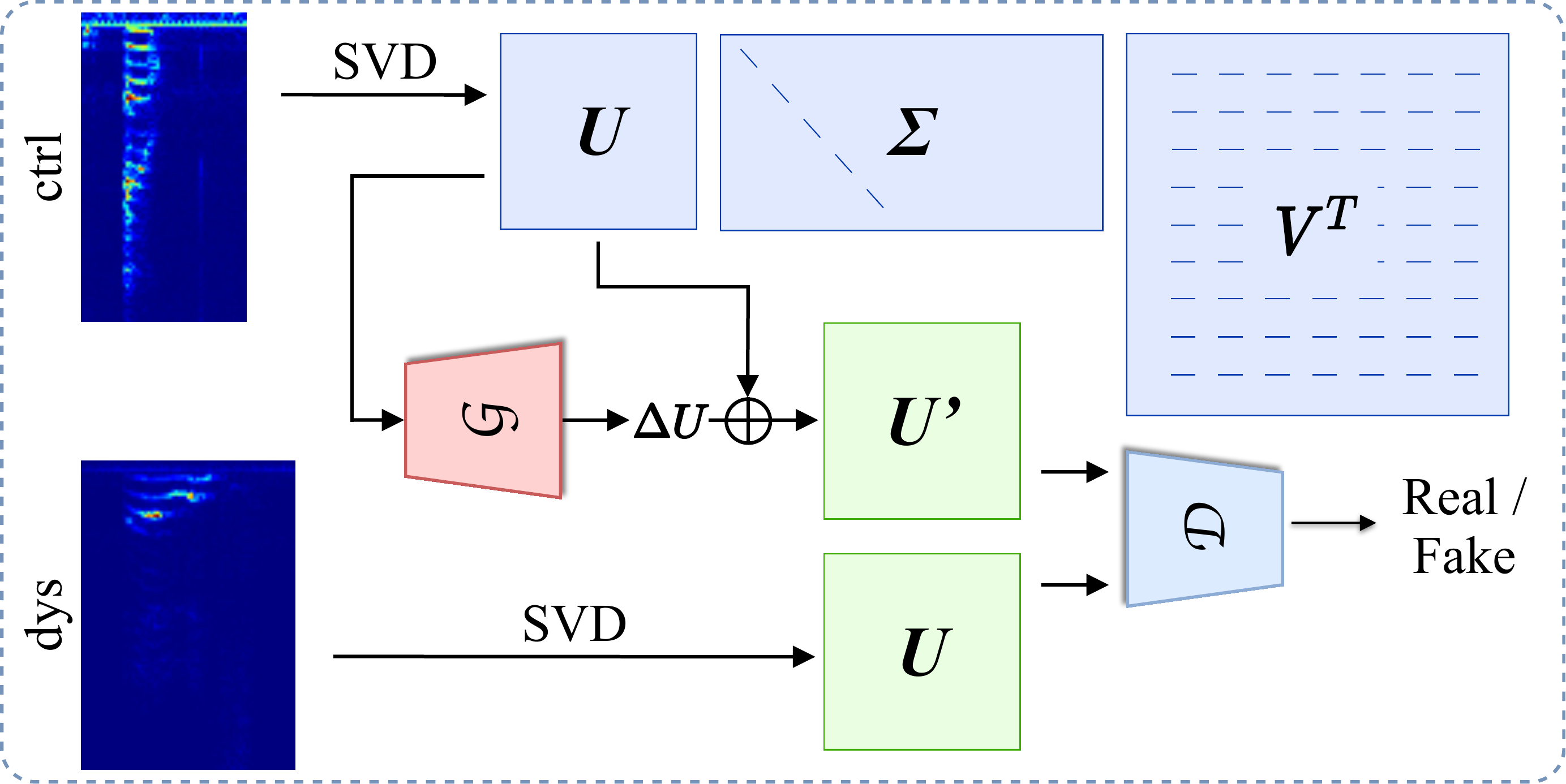}
    \label{fig:subfig:training_non_parallel}
}\hfill
\subfloat[] {
    \includegraphics[height=0.203\linewidth]{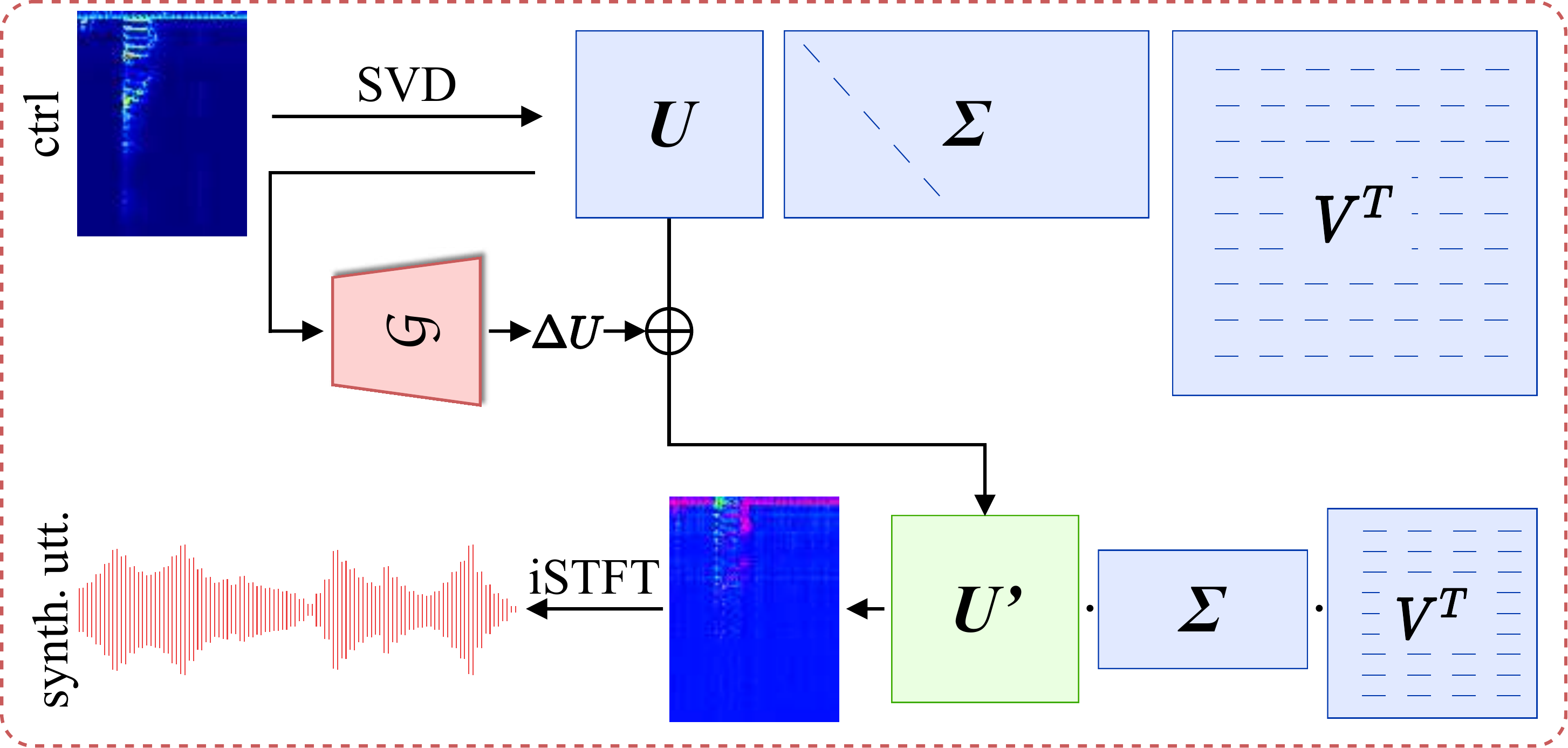}
    \label{fig:subfig:conversion_non_parallel}
}
\vspace*{-3mm}
\caption{Illustration of \textbf{(a)} DCGAN model training on parallel control and dysarthric utterances with modified duration and time alignment; \textbf{(b)} DCGAN based speaker-dependent (SD) dysarthric speech generation using SD speed perturbed normal speech; \textbf{(c)} Spectral basis GAN model training on SVD decomposed non-parallel control and dysarthric speech; and \textbf{(d)} Spectral basis GAN based SD dysarthric speech generation by re-composition of perturbed control speech derived spectral basis vectors with their temporal bases. }
\label{fig:training_non_parallel}
\vspace*{-7mm}
\end{figure*}

%% file: 2-SSL_models.tex
\vspace*{-1mm}
\subsection{Pre-trained Wav2vec2.0 Model}\label{wav2vec}
\vspace*{-1.5mm}
\textbf{Model architecture:} Wav2vec2.0 model consists of three components, including 1) a multi-layer CNN-based feature encoder which encodes raw speech $\mathcal{X}$ into continuous speech representations $z_t \in \mathcal{Z}$, 2) a transformer-based context network producing context representations $c_t \in C$ over the entire sequence of randomly masked feature encoder outputs and 3) a quantization module generating discrete speech units $q_t \in \mathcal{Q}$ as labels for self-supervised training.

\noindent\textbf{SSL pseudo-labels:} The Wav2vec2.0 model builds discrete unit $q_t$ as pseudo-label by quantization module for self-supervised pre-training.
After mapping the feature encoder output $z_t$ to the logit distribution $l \in \mathbb{R}^{G\times V}$, one entry is chosen from each of $G$ codebooks with $V$ entries by Gumbel-softmax re-parameterization.
The discrete unit $q_t$ is then obtained by applying a linear transformation to the concatenation of all $G$ chosen entries.

\noindent\textbf{SSL criterion:} The Wav2vec2.0 model is pre-trained via an interpolation between a contrastive task $\mathcal{L}_m$ and a diversity task $\mathcal{L}_d$. 
\vspace{-2mm}
\begin{equation}
\resizebox{0.915\linewidth}{!}{
\begin{math}
     \mathcal{L}_{w2v}=\underbrace{-\log \frac{\exp \left(\operatorname{sim}\left(c_t, q_t\right) / \kappa\right)}{\sum\limits_{\tilde{q} \sim Q_t} \exp\left(\operatorname{sim}\left(c_t, \tilde{q}\right) / \kappa\right)}}_{\text{\normalsize$\mathcal{L}_m$:} {\text{\normalsize ~Contrastive task}}} + \underbrace{\frac{1}{G V} \sum\limits^G_{g=1} \sum\limits_{v=1}^V \bar{l}_{g, v} \log \bar{l}_{g, v}}_{\text{\normalsize$\mathcal{L}_d$:}{\text{\normalsize ~Diversity task}}}
\end{math}
}
\label{eqn:wav2vec}
\vspace{-2mm}
\end{equation}
where $\operatorname{sim}(c_t, q_t)$ is the cosine similarity between the contextual representations produced before and after quantization. $\tilde{q}$ is a ``Distractor'' label. $\kappa$ is the non-negative temperature. 
The entropy-based diversity loss in the 2nd term ensures that Wav2vec2.0 utilizes all codebook entries equally. 
$\bar {l}_{g,v}$ is the average of logit distribution $l$ of $v$-th entry in $g$-th codebook across a mini-batch.

\noindent\textbf{Fine-tuning:} During fine-tuning, a randomly initialized linear layer is added on top of the context network to project representations into $C$ classes representing the target vocabulary, the output of which is optimized by Connectionist Temporal Classification (CTC) loss $\mathcal{L}_{CTC}$ on labeled speech data.

\vspace*{-3mm}
\subsection{Pre-trained HuBERT Model}
\vspace{-1.5mm}
The HuBERT pre-training process alternates between two steps: 1) a clustering step to create pseudo-labels and 2) a prediction step where the model produces labels at masked positions.

\noindent\textbf{Model architecture:} of the HuBERT model is similar to  Wav2vec2.0, including a feature encoder, a K-means quantization module and a transformer-based context network followed by a projection layer.

\noindent\textbf{SSL pseudo-labels:} The HuBERT model builds discrete speech unit $q_t$ as the pre-training pseudo-label via a separate k-means based clustering ensemble process.
A latent speech representation $s_t$ is quantized into discrete  units $q^{1}_{t}, q^{2}_t, \cdots, q^{N}_{t}$ by $N$ k-means featuring different codebook sizes. 
To improve the quality of pseudo-labels, embeddings from the intermediate layer of the context network at the $t$-th time step serve as latent speech representation $s_t$ to replace the initially selected MFCC features after the second clustering step.

\noindent\textbf{SSL criterion:} The cross-entropy loss guides the transformer-based context network to predict the discrete units of the masked continuous speech representations during pre-training. 
Such loss function $\mathcal{L}_m$ is only computed over the continuous speech representations $Z$ at masked time steps as
\vspace{-2.5mm}
\begin{equation}
    \mathcal{L}_m(Z, \{q^{(n)} \}_{N}, M) = \sum_{t \in M} \sum_n \log p^{(n)}\left(q_t^{(n)} \mid \tilde{Z}, t\right)
\vspace{-2.5mm}
\end{equation}
where $M$ represents the indices of all masked continuous speech representations, $\tilde{Z}$ represents the masked version of input sequence $Z$ and $p^{(n)}(q_t^{(n)} |\tilde{Z}, t)$ denotes the probability of the discrete speech units of the $t$-th frame assigned by the $n$-th K-means model. 

\noindent\textbf{Fine-tuning:} During fine-tuning, the projection layer is replaced by a Softmax layer and the model parameters are fine-tuned on labeled data using CTC loss with the CNN-based feature encoder frozen. 

\vspace*{-3mm}
\subsection{Speech Impairment Severity Based Multitask Fine-tuning}
\vspace{-1.5mm}
Motivated by \cite{geng23b_interspeech}, an additional speech impairment severity prediction task $\mathcal{L}_{\text{Seve}}$ is introduced to prevent overfitting to the limited amount of labeled UASpeech data during Wav2vec2.0 and HuBERT fine-tuning stage.
The total loss function of MTL is defined as
\vspace{-2.5mm}
\begin{equation}
\mathcal{L}_{M T L}=\beta_1 \cdot \mathcal{L}_{C T C}+\beta_2 \cdot \mathcal{L}_{\text {Seve }}
\vspace{-2.5mm}
\end{equation}
where $\beta_1 = \beta_2 = 0.5$ are empirically set hyper-parameters. Following \cite{geng23b_interspeech}, the CNN-based feature encoder is frozen when fine-tuning both Wav2vec2.0 and HuBERT models on UASpeech data.

%% file: 3-Conventional-DA.tex

\vspace{-1.5mm}
\subsection{Speed Perturbation Based Data Augmentation}
\label{subsec:sp}
\vspace{-1mm}
Speed perturbation resamples a given audio signal $x(t)$ in time domain \cite{ko15_interspeech}.
Given a perturbation factor $\alpha$, output audio segment is produced by applying $\alpha$ along the time axis as $y(t) = x (\alpha t)$.
This equation is equivalent to the following change in frequency domain as $X(f) \rightarrow \frac{1}{\alpha} X (\frac{1}{\alpha} f)$, where $X$ and $\frac{1}{\alpha} X (\frac{1}{\alpha} f)$ stand for the Fourier transform of $x(t)$ and $y(t)$ respectively.
Speed perturbation alters both audio duration and spectral envelope of the input audio.
A fixed set of speaker-independent (SI) perturbation factors $\{0.9, 1.0, 1.1\}$ is applied to disordered speech in common
with the data perturbation methods widely used in normal ASR tasks \cite{ko15_interspeech}.

\vspace{-2.5mm}
\subsection{Speaker Dependent Perturbation Based Augmentation}
\label{subsec:sd_speed}
\vspace{-1mm}
Due to the large mismatch in speaking rate between normal and disordered speakers, and the large variability among impaired speakers, speaker-dependent (SD) perturbation factors were applied when modifying normal speech to disordered speech.
These SD perturbation factors were obtained using phonetic analysis described in \cite{xiong2019phonetic}.
For each control speaker $C_i$ and dysarthric speaker $D_j$, we performed force alignment and calculated the average phoneme duration $l_{C_i}$ and $l_{D_j}$.
The average phoneme duration of all control speakers $l_{\overline{C}}$ is taken as the reference to compute
the speaker-dependent perturbation factor $F_{D_j}$ as $\frac{l_{\overline{C}}}{l_{D_j}}$, which was then used to modify normal speech to resemble that of dysarthric speaker $D_j$.

%% file: 4-Adversarial-DA.tex


\vspace{-0.5mm}
\subsection{DCGAN based Data Augmentation}
\label{subsec:adv_parallel}
\vspace{-1mm}

The SD speed perturbation of Sec. 3.2 only simulates the overall slower speaking rate and ``disordered like''  spectral properties of dysarthric speech. To capture more fine-grained spectro-temporal features of impaired speech, GAN-based data augmentation approaches can be used.
For the UASpeech dysarthric speech corpus that is based on parallel normal-impaired speech recordings, 
speaker-dependent deep convolutional GANs (DCGANs) are utilized to further transform SD speed perturbed healthy speakers' data to that of individual target impaired speakers of identical contents.

The architecture of DCGANs follows \cite{jin21_interspeech}. 
The Generator comprises 4 convolutional layers, each followed by a ReLU activation. 
The first three layers are of 8 kernels while the last one has 1 kernel.
All of the kernels in the Generator have a kernel size of 
$3 \times 3$ and stride of $1 \times 1$.
Replicate Padding is applied to ensure consistency in dimensions of the input and output of each convolutional layer.
The Discriminator consists of 4 convolutional layers containing $8$, $16$, $32$, and $64$ kernels respectively. All kernels have kernel size and stride of $2 \times 2$. 
A linear layer followed by a Sigmoid function is appended for binary classification. 
The output of the last convolutional layer is padded zero and flattened as the input of the linear layer.

Prior to DCGAN training, pairs of normal and dysarthric speech utterances of identical word contents are formed. 
In order to facilitate a frame-by-frame comparison between the GAN-transformed normal speech spectrogram against the target impaired spectrogram, each normal utterance is speed perturbed to temporally match the paired impaired utterance as shown in Fig. \ref{fig:subfig:training}. 
This requires a scaling factor to be estimated for each normal and dysarthric speech segment pair using phonetic analysis similar to the procedure described in \cite{xiong2019phonetic}.
The resulting pairs of normal and dysarthric speech utterances now have the same duration before being used in DCGAN training.
The DCGAN training maximizes the binary classification accuracy on the target dysarthric speech, and minimizes that obtained on the GAN-transformed normal speech.
This is given by
\vspace{-1.5mm}
\begin{equation}
\resizebox{0.91\linewidth}{!}{
\begin{math}
    \begin{aligned}
    \mathop{min}\limits_{G_j}^{}&\ \mathop{max}\limits_{D_j}^{} V(D_j, G_j) \\
    &\ = \mathbb{E}_{\textbf{f}_{\textbf{D}}\sim p_{D_j}(\textbf{f})}[\log{(D_j(\textbf{f}_{\textbf{D}_\textbf{j}}))}] + \mathbb{E}_{\textbf{f}_{\textbf{C}} \sim p_{C}(\textbf{f})}[\log{(1-D_j(G_j(\textbf{f}_{\textbf{C}})))}]
    \end{aligned}
\end{math}
}
\vspace{-1.5mm}
\end{equation}
where $j$ is the index for target dysarthric speaker, $G_j$ and $D_j$ are Generator and Discriminator for dysarthric speaker $j$, $\textbf{f}_\textbf{C}$ and $\textbf{f}_{\textbf{D}_\textbf{j}}$ are the STFT features of paired control and dysarthric utterances.

\vspace{-2.5mm}
\subsection{Spectral basis GAN based Data Augmentation}
\label{subsec:adv_non_parallel}
\vspace{-1mm}
In contrast to the DCGAN approach of Sec. \ref{subsec:adv_parallel} designed for parallel data,  more general and flexible adversarial data augmentation based on Spectral basis GANs \cite{jin2022personalized} that do not require parallel normal-impaired speech is utilized. 
Singular Value Decomposition (SVD) decomposed spectral bases \cite{geng21b_interspeech} are used as the inputs instead of STFT features. The convolutional layers in the DCGAN model are 
also replaced by fully connected layers.

The detailed architecture of Spectral basis GANs follows \cite{jin2022personalized}. 
The Generator consists of three linear layers where the first two are followed by a Leaky ReLU with 0.2 negative slope and the last one is connected to a Tanh activation.
The Discriminator contains three linear layers of $1024$, $512$ and $256$ dimensions each before the Sigmoid layer for binary classification.
The output targets of the Generator serve as a perturbation vector, $\Delta \mathbf{U}$, to be added to the spectral bases that are derived from the input normal speech spectrogram, $\mathbf{U}$.

%% file: 5-Experiments.tex
\input{Table/tab1}

\vspace{-2mm}
\subsection{Task Description}
\vspace{-1.5mm}
The UASpeech corpus \cite{uaspeech2008} is the largest publicly available dataset containing 103h speech from 16 dysarthric speakers and 13 healthy control speakers. 
The speech data of every speaker is split into three blocks B1, B2 and B3, each containing the same 155 common words and a different set of 100 uncommon words.
Data from B2 of all 16 dysarthric speakers is treated as the test set, while B1 and B3 data of all speakers are treated as the training set. 
With force alignment and silence stripping performed via an HTK \cite{young2002htk} trained GMM-HMM system, the dataset produces a 30.6h training set (99195 utt.) and an 8.7h test set (26520 utt.). 
Data augmentation using speaker-independent speed perturbation (Sec. \ref{subsec:sp}) produces a 72h training set, while speaker-dependent speed perturbation could further expand the training set to 130h.
B2 data of the 13 control speakers are further folded in, producing a 37h unaugmented (122392 utt.) and a 173h augmented training set (538292 utt.).

\vspace*{-4mm}
\subsection{Experiment Setup}
\vspace*{-1.5mm}
The \textbf{hybrid LF-MMI factored TDNN} systems follow the Kaldi chain system setup, except i-Vector features excluded. 
The \textbf{graphemic end-to-end (E2E) Conformer} systems implemented via ESPnet \cite{watanabe2018espnet} take 40-dim filter-bank + $\Delta$ features as inputs.
The \textbf{SSL Wav2vec2.0 and HuBERT} systems pre-trained with 60k hours of Libri-light data and initially fine-tuned on 960h LibriSpeech data, before 
further cross-domain fine-tuning on UASpeech data, with or without speed perturbation or GAN-based data augmentation.

\vspace{-3.5mm}
\subsection{Result Analysis}
\vspace{-1.5mm}
\noindent
\textbf{Performance of data augmentation:}
Table \ref{tab:WAV2VEC} presents the performance of LHUC-SAT adapted hybrid TDNN, E2E Conformer, Wav2vec2.0 and HuBERT systems incorporated with different data augmentation approaches applied on the training set. 
Several trends can be observed:
\textbf{1)} GAN-based data augmentation approaches consistently outperform fine-tuned Wav2vec2.0 and HuBERT models using no data augmentation and speed perturbation (Sys. 18, 19 \textit{vs.} Sys. 15, 17; Sys. 23, 24 \textit{vs.} Sys. 20, 22) by up to \textbf{2.01\% and  0.96\% absolute} (\textbf{9.03\% and 4.63\% relative}) WER reduction (Sys. 18 \textit{vs.} 15 and Sys. 23 \textit{vs.} 22, last col.).
\textbf{2)} On TDNN systems, the use of B2 control data brings up to 4.59\% absolute (15.7\% relative) WER reduction (Sys. 5 \textit{vs.} 2, last col.). 
\textbf{3)} On E2E Conformer systems, the use of B2 control data produced 20.08\% absolute (40.65\% relative) WER reduction (Sys. 14 \textit{vs.} 11, last col.), showing much larger performance sensitivity to improved training data coverage than TDNN systems. The Spectral basis GAN-based approach also outperforms speaker-dependent speed perturbation approach by \textbf{4.42\% absolute} (\textbf{13.1\% relative}) WER reduction (Sys. 14 \textit{vs.} 12, last col.) when both expands the training data to 173 hours.
\textbf{4)} The speed GAN approach (Sys. 18, 23) marginally outperforms Spectral basis GAN based augmentation on fine-tuned Wav2vec2.0 and HuBERT models by {0.41\% absolute} ({2.0\% relative}) overall WER reduction (Sys. 18 \textit{vs.} 19, last col.), while requiring more specialized parallel normal-impaired speech data in GAN model training $^1$ \footnote{$^1$ Prior researches found that the gains from speaker-dependent speed perturbation (the first step of speed GAN) are limited, e.g. when operating on non-parallel healthy vs. elderly speech \cite{ye21development}.}. 


\noindent
\textbf{Performance of system combination:} 
Sys. 25-27 in Table \ref{tab:WAV2VEC} present the performance of system combination using systems trained with speed (S), speed GAN (SG) and Spectral basis GAN (SBG) via two-pass rescoring \cite{cui2022two}. Larger WER reductions were obtained after combination by exploiting system diversity. For example, combining speed GAN data augmented systems produced an average WER of \textbf{16.69\%} (Sys. 26, last col.).
Sys. 28 in Table \ref{tab:WAV2VEC} further combines Sys. 25-27 and gives an overall WER of \textbf{16.53\%} (\textbf{46.47\% on the very low subset}) on the test set of 16 dysarthric speakers. To the best of our knowledge, this is the lowest WER reported on the UASpeech dataset, when compared with the most recent published results on the same task in Table 2.
\vspace{-2mm}
\input{Table/tab3}

%% file: Table/tab1.tex
\begin{table}[!t]
    \caption{Performance of LHUC-SAT adapted hybrid TDNN \cite{geng23b_interspeech}, end-to-end Conformer (CONF.), SSL Wav2vec2.0 (W2V.), SSL HuBERT systems and system combination (Sys. Comb.) via two-pass rescoring. ``MTL'' represents multi-task learning. ``B1\&3'' and ``B2'' in the ``Control'' col. represent control utterances from Block 1 \& 3 and Block 2. ``S'', ``SG'' and ``SBG'' denote speaker-dependent speed perturbation, speed GAN, and Spectral basis GAN. ``S'' in the ``Dys'' col. represents the speaker-independent speed perturbation. ``2x'' and ``5x'' refer to the amount of augmented data. ``Paral. Data'' stands for whether the data augmentation technique requires parallel data. ``VL'', ``L'', ``M'' and ``H'' denote intelligibility subgroups (Very Low/Low/Medium/High). ``$\dag$'', ``$\star$'' and ``$\diamond$'' denotes statistically significant (MAPSSWE \cite{gillick1989some}, $\alpha = 0.05$) improvement is obtained over the corresponding baselines (Sys. 1, 8, 16, 21 for ``$\dag$'', Sys. 5, 12, 17, 22 for ``$\star$'' and Sys.23 for ``$\diamond$'').  For Sys. 25-28, ``+'' represents score interpolation, while ``X$\rightarrow$Y'' denotes two pass rescoring \cite{cui2022two} the $N$-best ($N=100$) outputs of system X by system Y.
    }
    \vspace{-2mm}
    \label{tab:WAV2VEC}
    \centering
    \vspace{-0.15cm}
    \setlength{\abovecaptionskip}{0.05cm}
    \renewcommand\arraystretch{1.1}
    \renewcommand\tabcolsep{1.0pt}
     \resizebox{0.99\linewidth}{!}{
     \begin{tabular}{c|c|p{0.4cm}<{\centering}|p{0.4cm}<{\centering}|p{0.6cm}<{\centering}|p{0.4cm}<{\centering}|p{0.4cm}<{\centering}|p{0.6cm}<{\centering}|c|c|c|cccc|c}
        \hline\hline
        \multirow{4}{*}{Sys.} &
        \multirow{4}{*}{\tabincell{c}{Model}} &
        \multicolumn{7}{c|}{Data Augmentation} &
        \multirow{4}{*}{\tabincell{c}{Paral.\\Data}} &
        \multirow{4}{*}{\#Hrs.} &
        \multicolumn{5}{c}{Word Error Rate \%} \\
        \cline{3-9} \cline{12-16} 
         & & \multicolumn{6}{c|}{Control}  & \multirow{1}{*}{Dys} & & &\multicolumn{4}{c|}{\multirow{2}{*}{Intelligibility Subgroup}} & \multirow{3}{*}{All} \\
         \cline{3-9}
         & & \multicolumn{3}{c|}{B1\&3} & \multicolumn{3}{c|}{B2} & B1\&3 & & & & & & \\
         \cline{3-9} \cline{12-15}
         & & S & SG & SBG & S & SG & SBG & S & & & VL & L & M & H & \\
        \hline\hline
        1 & \multirow{7}{*}{\tabincell{c}{TDNN\\(LHUC\\-SAT)}} & - & - & - & - & - & - & \multirow{7}{*}{2x}& \xmark &72 & 63.80 & 28.42 & 18.82 & 8.13 & 27.23\\
        \cline{3-8}\cline{10-16}
        2 & & 2x & - & - & \multicolumn{3}{c|}{\multirow{3}{*}{excluded}} & & \xmark  & \multirow{3}{*}{130} & 61.64 & 30.34 & 20.51 & 13.14 & 29.23\\
        \cline{10-10}
        3 & & - & 2x & - & \multicolumn{3}{c|}{\multirow{3}{*}{}} & & \cmark & & 57.84 & 29.66 & 20.45 & 12.89 & 28.09    \\
        \cline{10-10}
        4 & & - & - & 2x & \multicolumn{3}{c|}{\multirow{3}{*}{}} & & \xmark & & 61.18 & 30.34 & 20.49 & 13.50 & 29.30\\
        \cline{3-8}\cline{10-16}
        5 &  & 2x & - & - & 5x & - & - & & \xmark & \multirow{3}{*}{173} & 61.62$^\dag$ & 24.56$^\dag$ & 15.82$^\dag$ & 6.50$^\dag$   & 24.64$^\dag$ \\
        \cline{10-10}
        6 & & - & 2x & - & - & 5x & - & & \cmark & & 57.34$^{\dag\star}$ & 24.52$^\dag$ & 15.86$^\dag$ & 6.23$^{\dag\star}$  & 23.64$^{\dag\star}$ \\
        \cline{10-10}
        7 & & - & - & 2x & - & - & 5x & &\xmark & & 61.03$^\dag$ & 24.65$^\dag$ & 16.02$^\dag$& 6.23$^{\dag\star}$  & 24.49$^\dag$ \\
        \hline\hline
        8 & \multirow{7}{*}{\tabincell{c}{CONF.\\(enc8+\\dec4)}} & - & - & - & - & - & - & \multirow{7}{*}{2x} & \xmark & 72 & 66.41 & 42.87 & 33.00 & 10.27 & 34.98 \\ 
        \cline{3-8}\cline{10-16}
        9 & & 2x & - & - & \multicolumn{3}{c|}{\multirow{3}{*}{excluded}} & &  \xmark & \multirow{3}{*}{130} & 66.06 & 48.40  & 46.35 & 41.60  & 49.45 \\
        \cline{10-10}
        10 & & - & 2x & - & \multicolumn{3}{c|}{\multirow{3}{*}{}} & & \cmark & & 66.34 & 48.43 & 46.23 & 41.59 & 49.49 \\
        \cline{10-10}
        11 & & - & - & 2x & \multicolumn{3}{c|}{\multirow{3}{*}{}} & & \xmark & & 66.54 & 47.71 & 46.27 & 41.73 & 49.40 \\
        \cline{3-8}\cline{10-16}
        12 &  & 2x & - & - & 5x & - & - & & \xmark  & \multirow{3}{*}{173} & 66.57 & 41.33$^\dag$ & 31.72$^\dag$ & 8.40$^\dag$  & 33.74$^\dag$ \\
        \cline{10-10}
        13 & & - & 2x & - & - & 5x & - & & \cmark && 66.14 & 36.47$^{\dag\star}$ & 24.37$^{\dag\star}$ & 5.40$^{\dag\star}$  & 29.96$^{\dag\star}$ \\
        \cline{10-10}
        14 & & - & - & 2x & - & - & 5x & &\xmark & & 66.18 & 34.00$^{\dag\star}$  & 23.92$^{\dag\star}$ & 5.65$^{\dag\star}$  & 29.32$^{\dag\star}$ \\
        \hline\hline
        15 & \multirow{5}{*}{\tabincell{c}{W2V.\\(MTL)}} & - & - & - & - & - & - & - & \xmark & 37  & 59.82 & 24.62 & 11.53 & 2.95 & 22.25 \\
        \cline{3-9}\cline{10-16}
        16 & & - & - & - & - & - & - & \multirow{4}{*}{2x} & \xmark & 72 & 57.79 & 23.65 & 11.04 & 2.76 & 21.41 \\
        \cline{3-8}\cline{10-16}
        17 & & 2x & - & - & 5x & - & - & & \xmark & \multirow{3}{*}{173} & 57.36 & 21.68$^\dag$ & 10.25$^\dag$  & 2.88  & 20.71$^\dag$ \\
        \cline{10-10}
        18 & & - & 2x & - & - & 5x & - & & \cmark & & 55.83$^{\dag\star}$ & 21.66$^\dag$ & 10.06$^\dag$ & 2.63$^\star$  & 20.24$^{\dag\star}$ \\
        \cline{10-10}
        19 & & - & - & 2x & - & - & 5x & &\xmark & & 56.49$^\dag$ & 22.09$^\dag$ & 10.57 & 2.77 & 20.65$^\dag$\\
        \hline\hline
        20 & \multirow{5}{*}{\tabincell{c}{Hu\\-BERT\\(MTL)}} & - & - & - & - & - & - & - & \xmark & 37 & 55.60  & 23.31 & 12.45 & 2.63 & 21.10 \\
        \cline{3-9}\cline{10-16}
        21 & & - & - & - & - & - & - & \multirow{4}{*}{2x} & \xmark & 72 & 58.36 & 24.10 & 13.73 & 2.30 & 22.01 \\
        \cline{3-8}\cline{10-16}
        22 & & 2x & - & - & 5x & - & - & & \xmark & \multirow{3}{*}{173} & 56.88$^\dag$ & 21.60$^\dag$ & 11.41$^\dag$  &  2.67 & 20.73$^\dag$ \\
        \cline{10-10}
        23 & & - & 2x & - & - & 5x & - & & \cmark & & 53.99$^{\dag\star}$ & 21.50$^\dag$ & 9.84$^{\dag\star}$ & 2.60  & 19.77$^{\dag\star}$ \\
        \cline{10-10}
        24 & & - & - & 2x & - & - & 5x & &\xmark & & 55.35$^{\dag\star}$ & 20.99$^\dag$ & 10.04$^{\dag\star}$ & 2.68 & 19.99$^{\dag\star}$ \\
        \hline\hline
        25 & \multirow{4}{*}{\tabincell{c}{Sys.\\Comb.}} & \multicolumn{9}{c|}{Sys. 5 + (5$\rightarrow$12) + (5$\rightarrow$17)  + (5$\rightarrow$22)} & 48.84$^\diamond$ & 16.73$^\diamond$ & 7.29$^\diamond$ & 3.08 & 17.12$^\diamond$ \\
        \cline{3-16}
        26 & & \multicolumn{9}{c|}{Sys. 6 + (6$\rightarrow$13) + (6$\rightarrow$18) + (6$\rightarrow$23)} & 46.90$^\diamond$ & 17.34$^\diamond$ & 6.71$^\diamond$ & 2.91 & 16.69$^\diamond$ \\
        \cline{3-16}
        27 & & \multicolumn{9}{c|}{Sys. 7 + (7$\rightarrow$14) + (7$\rightarrow$19) + (7$\rightarrow$24)} & 47.99$^\diamond$ & 17.08$^\diamond$ & 7.80$^\diamond$ & 2.85 & 17.04$^\diamond$ \\
        \cline{3-16}
        28 & & \multicolumn{9}{c|}{Sys. 25 + 26 + 27} & 46.47$^\diamond$ & 16.76$^\diamond$ & 7.18$^\diamond$ & 2.89 & 16.53$^\diamond$ \\
        \hline\hline
    \end{tabular}}
    \vspace{-7.5mm}
\end{table}

%% file: Table/tab3.tex
\begin{table}[!ht]
    \vspace{-0.5mm}
    \centering
    \caption{A comparison between published systems on UASpeech and ours. Naming conventions follow the one adopted in Table \ref{tab:WAV2VEC}.}
    \label{tab:comparison}
    \vspace{-0.35cm}
    \renewcommand\tabcolsep{1pt}
    \resizebox{0.92\linewidth}{!}{
    \begin{tabular}{c|c|c|c}
    \hline
    \hline
    Performance (WER\%) of Systems Published on UASpeech& VL & L & All \\
    \hline
CUHK-2020 DNN + DA + LHUC-SAT \cite{geng2020investigation} & 62.44 & 27.55 & 26.37 \\
CUHK-2021 DNN + DCGAN + LHUC-SAT \cite{jin21_interspeech} & 61.42 & 27.37 & 25.89 \\
Nagoya Univ.-2022 WavLM \cite{lester2022investigatiing} & 71.50 & 50.00 & 51.80 \\
Brno Univ.-2022 Wav2vec2 + SAT (15 spkr) \cite{baskar22b_interspeech} & 57.72 & 22.46 & 22.83\\
FAU-2022 Cross-lingual XLRS + Conformer \cite{hernandez22_interspeech} & 62.00 & 28.60 & 26.10 \\
BTBU-2022 Multi-stage AVHuBERT \cite{AVSSL} & 63.98 & 30.77 & -\\
CUHK-2022 DNN + Data Aug. + SBE Adapt + LHUC-SAT \cite{mengzhe22SVD} & 59.30 & 26.25 & 25.05\\
CUHK-2023 Kaldi TDNN + VAE-GAN + LHUC-SAT \cite{jin23vae} & 57.31 & 28.53 & 27.78 \\
JHU-2023 DuTa-VC (Diffusion) + Conformer \cite{wang23qa_interspeech} & 63.70 & 27.70 & 27.90\\
CUHK-2023 TDNN + Wav2vec2.0 feat. + Sys. Combination \cite{hu2023exploring} & 53.12 & 25.03 & 22.56 \\
CUHK-2023 DNN + Wav2vec2.0 + Sys. Combination \cite{geng23b_interspeech} & 51.25 & 17.41 & 17.82 \\
\hline
\makecell[c]{\textbf{Ours, Wav2vec2.0/HuBERT + GAN Data Aug. }\\ \textbf{+ Sys. Comb (Sys. 28, Table \ref{tab:WAV2VEC})}} & \textbf{46.47} & \textbf{16.76} & \textbf{16.53}\\
\hline
\hline
\end{tabular}
}
\vspace{-6mm}
\end{table}

%% file: 6-Conclusions.tex
This paper proposes two GAN-based data augmentation approaches for robust domain fine-tuning of pre-trained Wav2vec2.0 and HuBERT models for dysarthric speech recognition. 
Experiments conducted on the UASpeech corpus suggest such adversarial approaches significantly improve their generalization performance.
Future research will focus on improving adversarial data augmentation to model richer spectro-temporal characteristics of dysarthric speech.